\documentclass[12pt,a4paper]{article}
\usepackage{amssymb,amsmath,graphicx,subcaption,hyperref,cite}
\usepackage{epstopdf}
\usepackage{tabularx}
\usepackage[utf8]{inputenc}
\usepackage[english]{babel}

\begin{document}
	
\begin{center}
	{\Large\bf Competitive metastable behaviours of surface and bulk in Ising ferromagnet }
\end{center}
\vskip 1 cm
\begin{center} %
	Moumita Naskar$^1$ and Muktish Acharyya$^{2,*}$
	
	\textit{Department of Physics, Presidency University,} 
	
	\textit{86/1 College Street, Kolkata-700073, India} 
	\vskip 0.2 cm
	\textit{Email$^1$:naskar.moumita18@gmail.com}
	
	\textit{Email$^2$:muktish.physics@presiuniv.ac.in}
\end{center}
\vspace {1.0 cm}
\noindent {\bf{Abstract:}} The reversal of magnetisation has been studied in a three dimensional Ising ferromagnet by Monte Carlo simulation with Metropolis single spin flip algorithm using random updating scheme. The outer layers are considered as surface. The surface interacts with core with a relative ferromagnetic interaction strength. Depending on the relative interaction strength, the time of reversal of the surface was found to be different from that of the bulk. For weaker relative strength, surface reversal was found to be faster than that of bulk and vice versa for stronger relative interaction strength. A critical value ($R_c$) of relative interaction strength provides same time of reversal of surface and bulk. This critical relative interaction strength was found to be a function of the temperature ($T$) and applied magnetic field ($h$). The scaling relation $R_c \sim h^{-\beta}f(Th^{\alpha})$, where, $\alpha=0.23\pm0.01$ and $\beta = -0.06\pm0.01$, has been proposed, numerically by the method of data collapse. The metastable volume fractions, for both surface and bulk, were found to follow the Avrami's law. The critical relative interaction strength ($R_c$) has been observed to decrease in an exponential ($e^{bL^{-1.5}})$ fashion with the system size ($L$).

\vskip 2cm
\noindent $^*$Communicating author: Muktish Acharyya (E-mail:muktish.physics@presiuniv.ac.in)
\newpage
\section{Introduction}
\vskip 0.5 cm

The mechanisms of the magnetic recording and the data storage are extremely important technological\cite{techno} aspects of the modern civilization.
The magnetisation switching (or reversal) plays the key role for
these mechanisms. As a result, the experimental and theoretical understanding of this reversal phenomenon became a challenging field of modern research. Starting from the phenomenological classical nucleation theory\cite{becker, gunton}, the computer
simulational studies\cite{mastauffer , rikvold1 , bkc , ma2010 , ma2014} have shown its importance in this field. The growth of the structure of nucleating cluster can be investigated by numerical simulational study. The rate of growth and decay of cluster has been studied\cite{vehkamaki} as function of applied external magnetic field and the temperature of the system.

Some recent interesting simulational studies, have drawn much attention of the researchers in this context. The relaxation of
Ising ferromagnet, after a sudden reversal of the direction of applied external magnetic field has been studied \cite{binder1}.
An interesting crossover from coherent rotation to the nucleation has been reported\cite{hinzke} in classical anisotropic Heisenberg ferromagnet. The heat assisted reversal has been studied \cite{rikvold2} by Monte Carlo simulation of ultrathin films. The effects of randomly quenched disorder
(random field) on the reversal have been reported \cite{moumita1} recently in Ising ferromagnets. The effects of gradient of the field \cite{abyaya} and the gradient of the temperature \cite{ranajay} in the reversal of Ising ferromagnet, have also been investigated by Monte Carlo (MC) simulation. Recently, the roles of anisotropy \cite{moumita2}and its
gradient \cite{moumita3} have been explored in the Blume-Capel (BC)
ferromagnets. The dynamical phase transition has been studied 
\cite{erol}in the
site diluted BC model. The universality, arising from disorder, was
also reported \cite{fytas} recently in the random bond Blume–Capel model. The mixed spin (S = 1, S = 1/2) Blume–Capel model
was studied \cite{selke} by Monte Carlo simulation and the absence
of tricritical point was noticed in two dimensions. Very recently, the metastable behaviours of general spin-s
Blume-Capel ferromagnet have been investigated \cite{mmen} by MC simulation.

The studies on the reversal of magnetisation discussed above, are mainly of homogeneous type. The strength of the interaction is mainly uniform over the space. Moreover, 
the reversal of magnetisation has not yet been investigated as a competition between the reversal of the surface and that of the bulk. It is not yet known, whether the reversals of magnetisations occur in the bulk and in the surface simultaneously or not. In this regard, it seems to be necessary to conduct an investigation of such a competitive (if any) behaviours of the magnetisation reversal of bulk and surface, at least by MC simulation in simple Ising ferromagnets. 

In this article, we address this issue. What will happen
in a bulk sample exposed to the open boundary condition? Do the surface and bulk show the same time of magnetisation reversal? 

This manuscript is organized as follows: in the next section (section-2) we have described the model and the methods of
the simulation. The numerical results are reported in section-3.
The paper ends with a few concluding remarks given in section-4.


\section {Description of the Model and Simulation Scheme}
\vskip 0.5 cm

\begin{figure}[h!]
	\centering
	\begin{subfigure}{0.49\textwidth}
	\includegraphics[angle=0,width=1.2\textwidth]{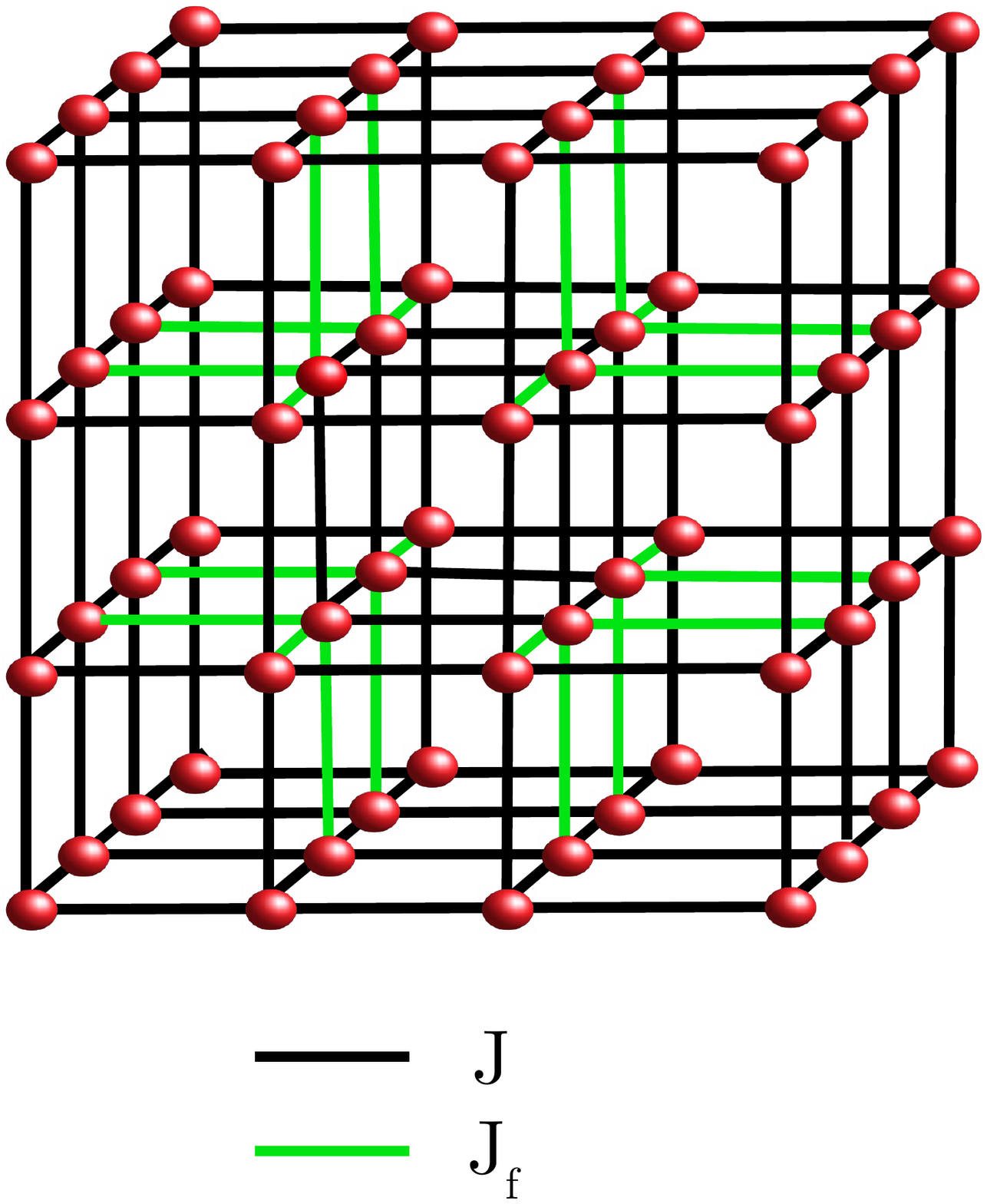}
	\subcaption{}
    \end{subfigure}
	\begin{subfigure}{0.49\textwidth}
	\includegraphics[angle=0,width=1.2\textwidth]{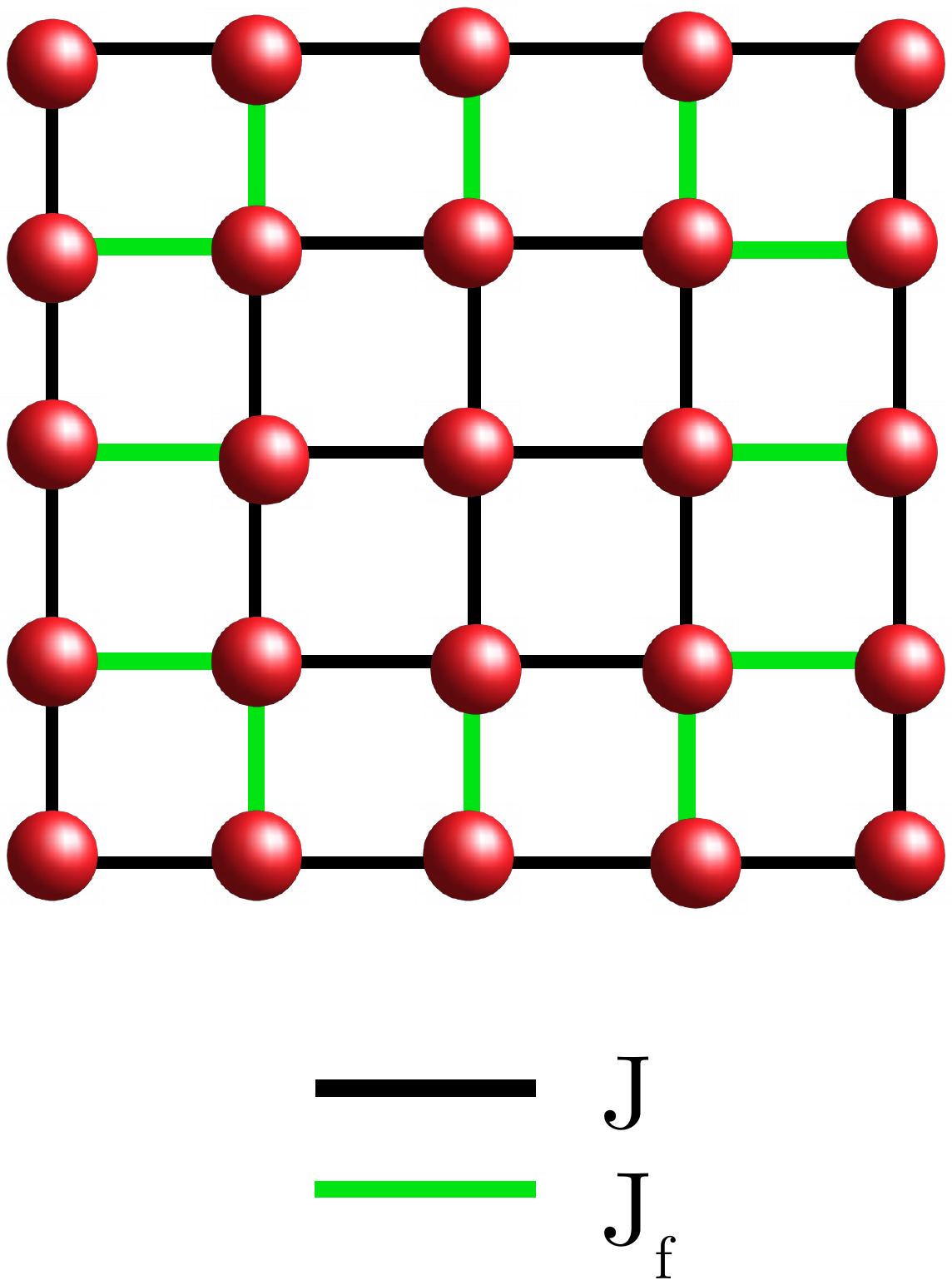}
	\subcaption{}
	\end{subfigure}
	\caption{(a) Schematic of a $4 \times 4 \times 4$ cubic lattice (b) Schematic of {\it two dimensional projection} of a $5 \times 5 \times 5$ cubic lattice. Interaction strengths $J$ and $J_f$ are indicated by black and green lines respectively.}
	\label{schematic}
\end{figure}

In order to study a comparative metastable behaviour of surface and bulk, the three dimensional ferromagnetic Ising system has been described here by the Hamiltonian,
\begin{equation}
	H=-\sum_{<i,j>} J\:S_i^z S_j^z -\sum_{<l,k>} J\: \sigma_l^z \sigma_k^z -\sum_{<p,q>} J_f \: S_p^z \sigma_q^z - h \sum_{i}(S_i^z+\sigma_i^z)
\end{equation}	
where $S_i^z$ (spin at i-th site of the core) and $\sigma_l^z$
(spin at l-th site of the surface) are the  Ising spins (can access two discrete values, i.e., +1 and -1 only). First term represents the contribution coming from the nearest neighbour interaction between the spin pairs within the core. The second term considers the contribution coming from the nearest neighbour interaction between pair of spins within the surface only. The third term captures the contribution coming from the interaction between core and surface. The last term represents the interaction with the applied external magnetic field ($h$). The nearest neighbour interaction strength between the core and the surface is taken as $J_f$ (green lines in Fig-\ref{schematic}) and all other types of (spin-spin) interaction are taken as $J$ (black lines in Fig-\ref{schematic}). Both $J$ and $J_f$ are ferromagnetic ($> 0$). In Fig-\ref{schematic}a, schematic of the lattice of size $L=4$ is presented. For convenience, a two dimensional projection of  the lattice of size $L=5$ is also provided (Fig-\ref{schematic}b). It may be mentioned here that
the relative interaction strength is defined as $R=J_f/J$.

In the present study, the surface consists of all the six outermost square layers of the cubic lattice. The system is kept in the open boundary conditions in all three directions. It may be worth mentioning here that such kind of interfacial interaction strength was used
\cite{pleiml1,pleiml2,berger} to study the surface critical behaviour of the nonequilibrium phase transition in driven kinetic Ising ferromagnets.

It should be clearly mentioned that, the whole system of $N=L^3$ number of spins (including the surface) is defined here as bulk. And the surface (say surface-i) consists of the six outermost square layers of the bulk (Fig-\ref{schematic}). The surface (surface-i), defined in this way, contains $N_s=L^3-(L-2)^3$ number of spins. For the convenience of discussion, let us denote the inner part of the bulk (excluding surface) as core. Definitely, the core contains $(L-2)^3$ number of spins. 

More precisely, there exist four categories of spins as far as the
coordination number is concerned. Namely, (i) each spin inside the core has six nearest neighbours, (ii) each of the eight spins at the
corners,  having three nearest neighbours, (iii) each spin on the twelve edges (excluding the corner sites), having four nearest neighbours only, (iv) rest defines the 
{\it restricted-surface}  (so called surface-ii), each spin on such surface having five nearest neighbours. According
to this classification, the core contains $(L-2)^3$ number of spins
and the surface contains $6\times(L-2)^2$ number of spins. It may be noted here that for large $L$ limit, the ratio of the
total number ($N_e= 12\times(L-2)$) of spins on the edges to the total number ($N'_s= 6\times(L-2)^2$) of spins on the {\it restricted-surface} would be vanishingly small. In this token,
the surface-i and surface-ii would yield the same result in the
context of reversal of magnetisation. In the present study, we
used surface-i.

Such a $ L\times L \times L $ three dimensional ferromagnetic Ising cubic lattice is considered for executing the following studies. Starting from perfectly ordered state where all the spins are pointing up $S_i^z \; {\rm and} \; \sigma_i^z =+1$ $\forall$ i, the lattice is updated using random updating scheme with \textbf{open boundary conditions} in all three directions. The
probability of the flipping of any spin, has been determined by Metropolis algorithm\cite{metrop}, 
$P(S_i^z\to-S_i^z) = Min(1,e^{- \frac{\Delta E}{k_B T}})$ (for spin in the core) or $P(\sigma_i^z\to-\sigma_i^z) = Min(1,e^{- \frac{\Delta E}{k_B T}})$ (for spin on the surface).
 Where $\Delta E$ is the change in energy due to spin flip. $k_B$ is the Boltzmann constant and $T$ is the temperature of the system (measured in the unit of $J/k_B$). The energies
are measured in the unit of $J$. The spin will flip, if the value of the probability $P$  exceeds a random number (uniformly distributed between 0 to 1).  The $N=L^3$ number of such random updates of lattice sites defines the unit of time in the present problem and called Monte Carlo Step per Site (MCSS). The total magnetisation of the bulk is determined by
\begin{equation}
m_b(t)=\frac{1}{N} \sum_{i}^{N} (S_i^z + \sigma_i^z)
\end{equation}
where $N=L^3$ is the total number of spins in the system. $S_i^z$ and $\sigma_i^z$ are the spins of the core and the surface respectively as described earlier. The magnetisation of the surface (only) is determined by
\begin{equation}
m_s(t)=\frac{1}{N_s} \sum_{ i } \sigma_i^z  
\end{equation}
where $\sigma_i^z$ are those spins which are situated on the surface (considered here) and $N_s= L^3-(L-2)^3$ is the total number of such spins.

\vskip 1 cm
\section{Simulational results}
\vskip 0.5 cm

Under the influence of a negative magnetic field, a ferromagnetic system below critical temperature $T<T_c$ sustains positive magnetisation for a certain period of time and then gradually acquires negative magnetisation with parallel alignment of the spins along the applied field. On the way of eventual decay of the metastable state to the true equilibrium state, time taken by the system at which the magnetisation just starts to alter its sign (or the system acquires zero magnetisation $m(t) \simeq 0$) is referred here as \textit{lifetime of metastable state} or the \textit{reversal time of magnetisation} ($\tau$). In this paper, our main goal is to explore a comparative study of the scale of the reversal time of surface and bulk. In particular, we will try to address the questions, \textit{do the surface and bulk consume same amount of time to leave the metastable state in presence of negative field? Or, does the surface reversal synchronize with the bulk reversal?}
 
\begin{figure}[h!]
	\centering
	\begin{subfigure}{0.49\textwidth}
		\includegraphics[angle=0,width=1.2\textwidth]{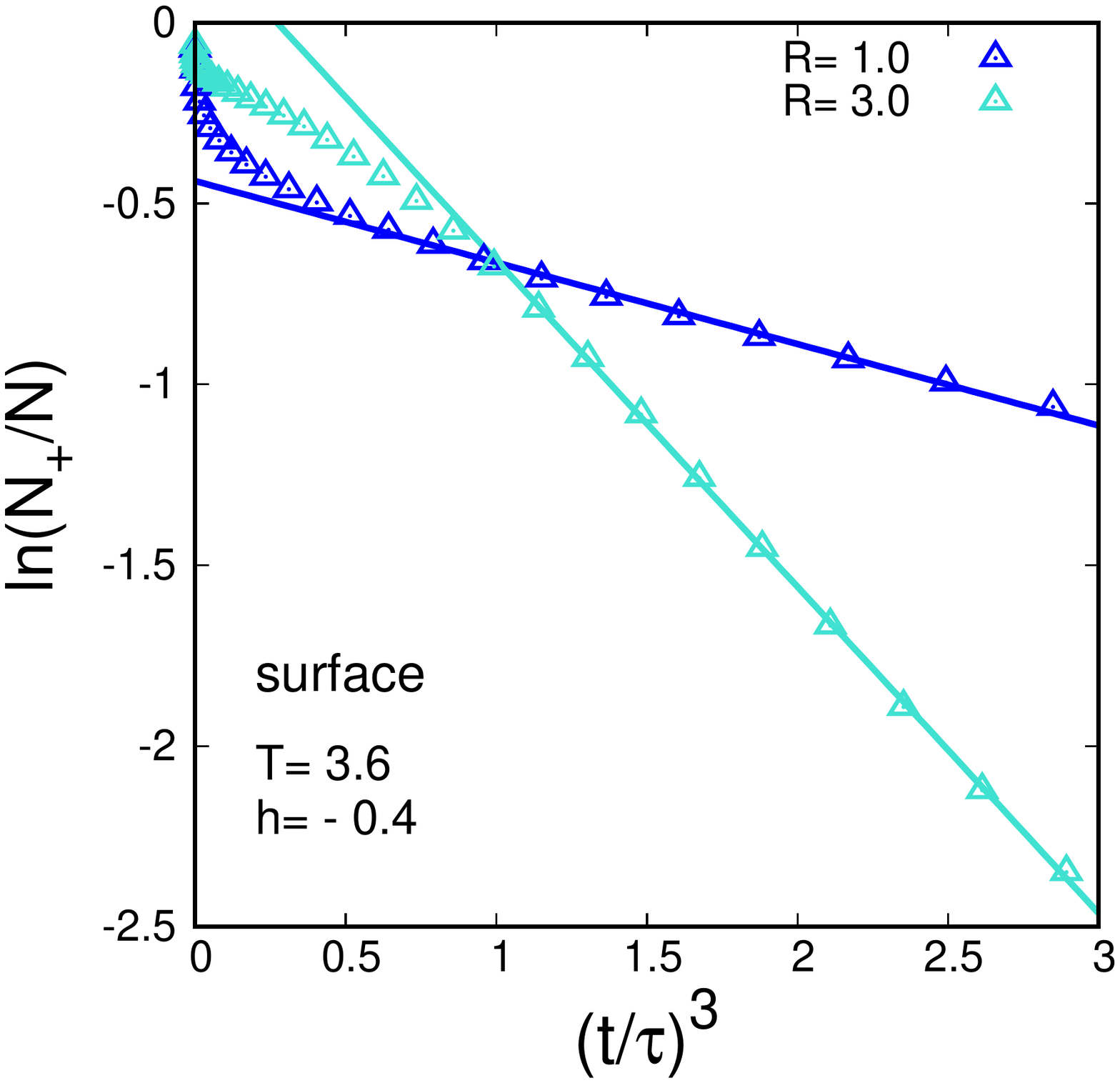}
		\subcaption{surface}
	\end{subfigure}
	\begin{subfigure}{0.49\textwidth}
		\includegraphics[angle=0,width=1.2\textwidth]{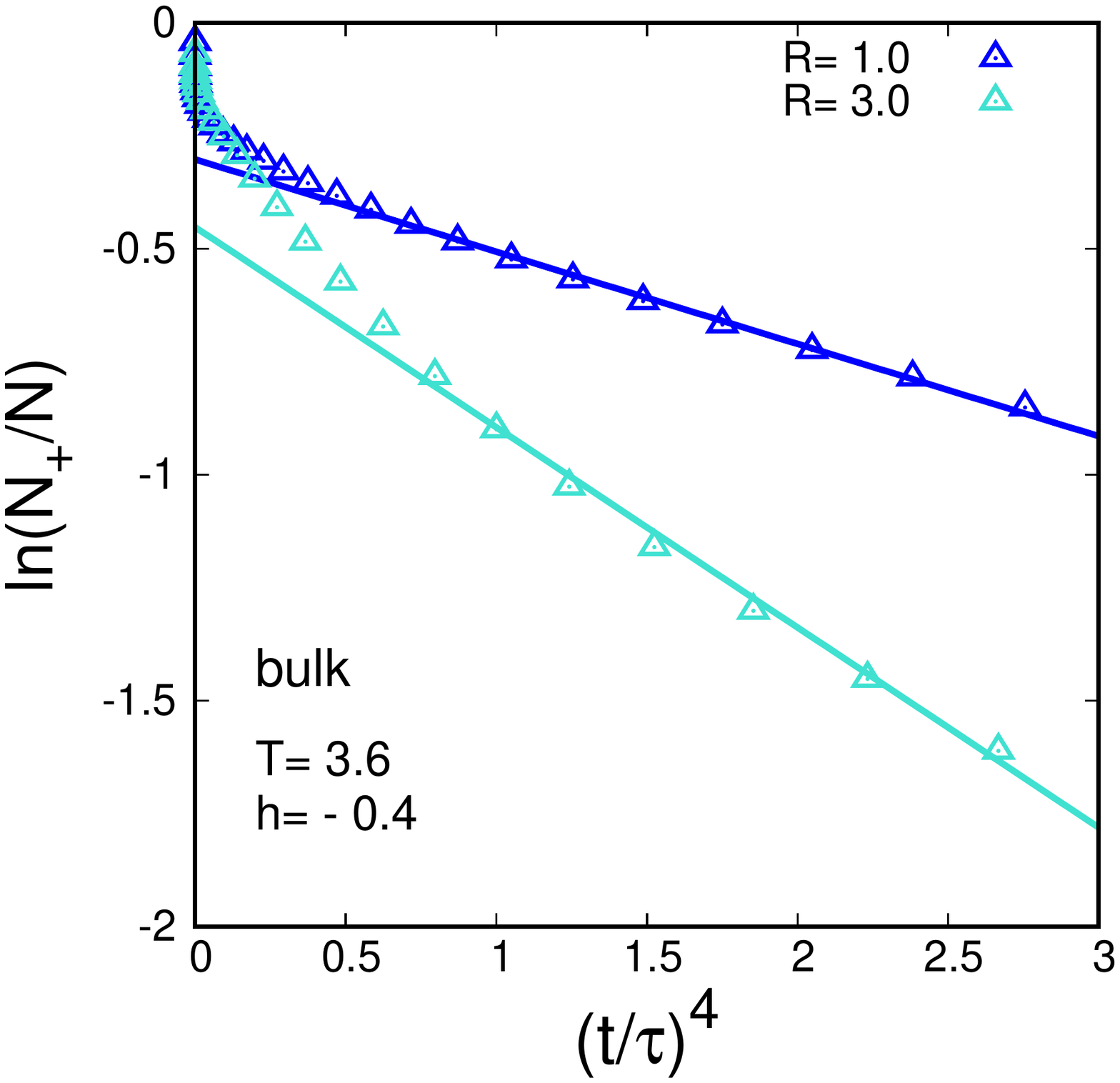}
		\subcaption{bulk}
	\end{subfigure}
	\caption{Evolution of logarithm of density of spin '+1' with time ($t^3$ for surface and $t^4$ for bulk) for two different relative interaction strengths $R= J_f/J = 1.0,3.0$. Data are averaged over 1000 samples. $\tau$ is the reversal time which is different for each plot. Temperature is fixed at $T=3.6$ (0.8 $T_c$) and the applied field is $h=-0.4$. $Here, L=32$}
	\label{avrami}
\end{figure}

 How does metastable volume fraction decay with time in the bulk and in the surface? What is metastable volume fraction? Since, the initial spin configuration is considered as all spins are up, the metastable volume fraction is defined as the relative abundance of up (+1) spin $(N_{+}/N)$. In this context, 
Avrami's law \cite{johnson,avrami,ramos} tells that the
$log(N_{+}/N) \sim -(t/{\tau})^{d+1}$, in $d$ dimensions. In order to study the evolution of metastable volume fraction, density of up spin is studied with time (below critical temperature) at the temperature $T=3.6 (0.8 \; T_c)$. For surface, logarithm of density of up spin on surface is studied with third $(2+1)$ power of time (Fig-\ref{avrami}(a)). For the relative interaction strength $R=1.0$ or $J_f=J=1.0$ the system obeys the Avrami's law for wider range of time which indicates that the surface behaves almost as a two dimensional system. If the relative interaction strength $R$ is increased ($R=3.0$) the range of validity of the law (Avrami) decreases. Initially, the volume fraction decays slowly but later ($t > \tau$) it decays faster. Similarly, for bulk, the logarithm of the relative abundance of up spin of the whole system is found to decay with fourth ($3+1$) power of time (indicating a three dimensional system)(Fig-\ref{avrami}(b)). For the relative interaction strength $R=1.0$ or $J_f=J=1.0$ the system follows the law for wider range of time. If the relative interaction strength is increased ($R=3.0$), initially the volume fraction decays faster but later ($t > \tau$) it decays with slower rate.

\begin{figure}[h!]
	\centering
\begin{subfigure}{0.49\textwidth}
	\includegraphics[angle=0,width=1.2\textwidth]{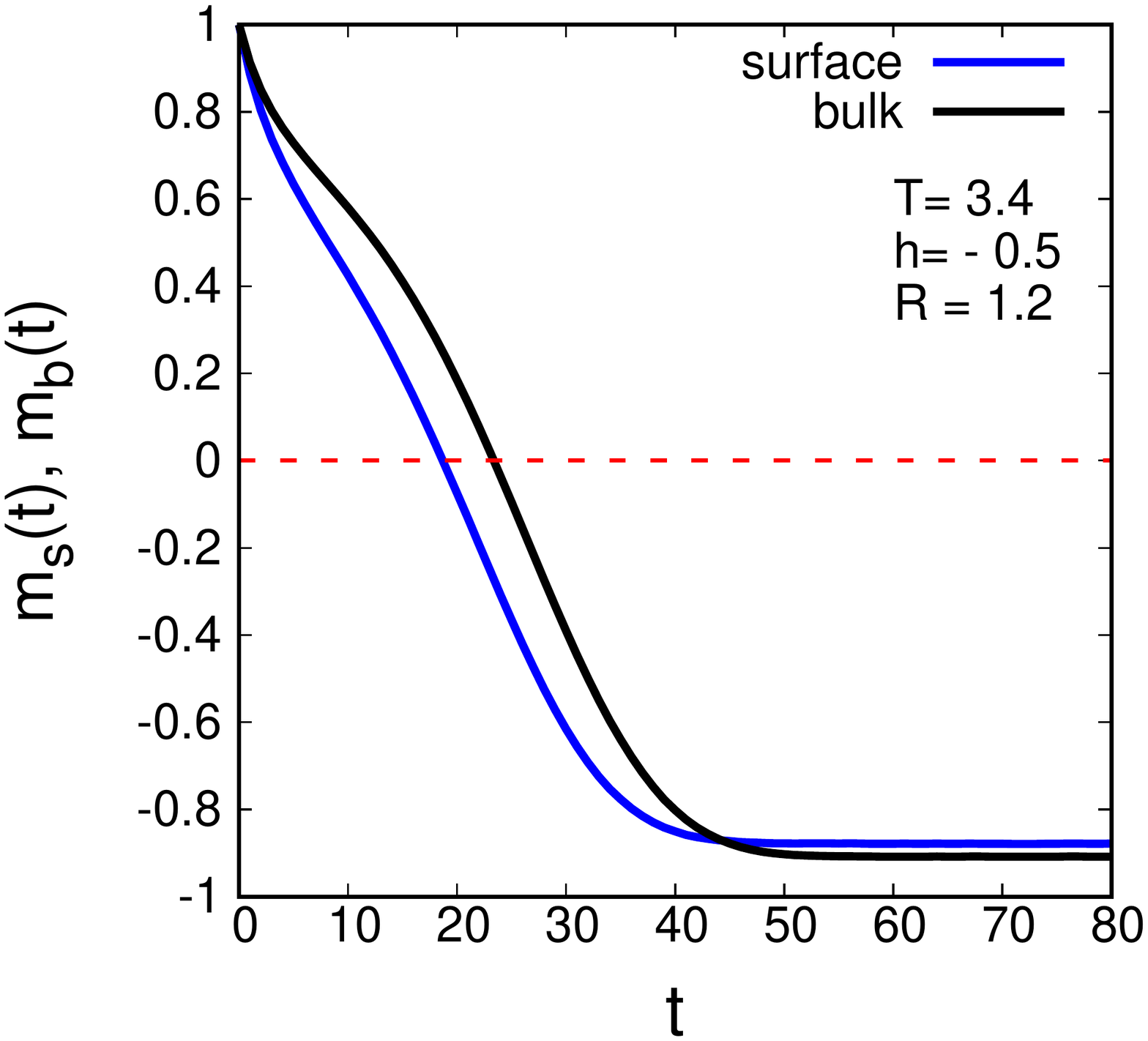}
	\subcaption{}
\end{subfigure}
	\begin{subfigure}{0.49\textwidth}
		\includegraphics[angle=0,width=1.2\textwidth]{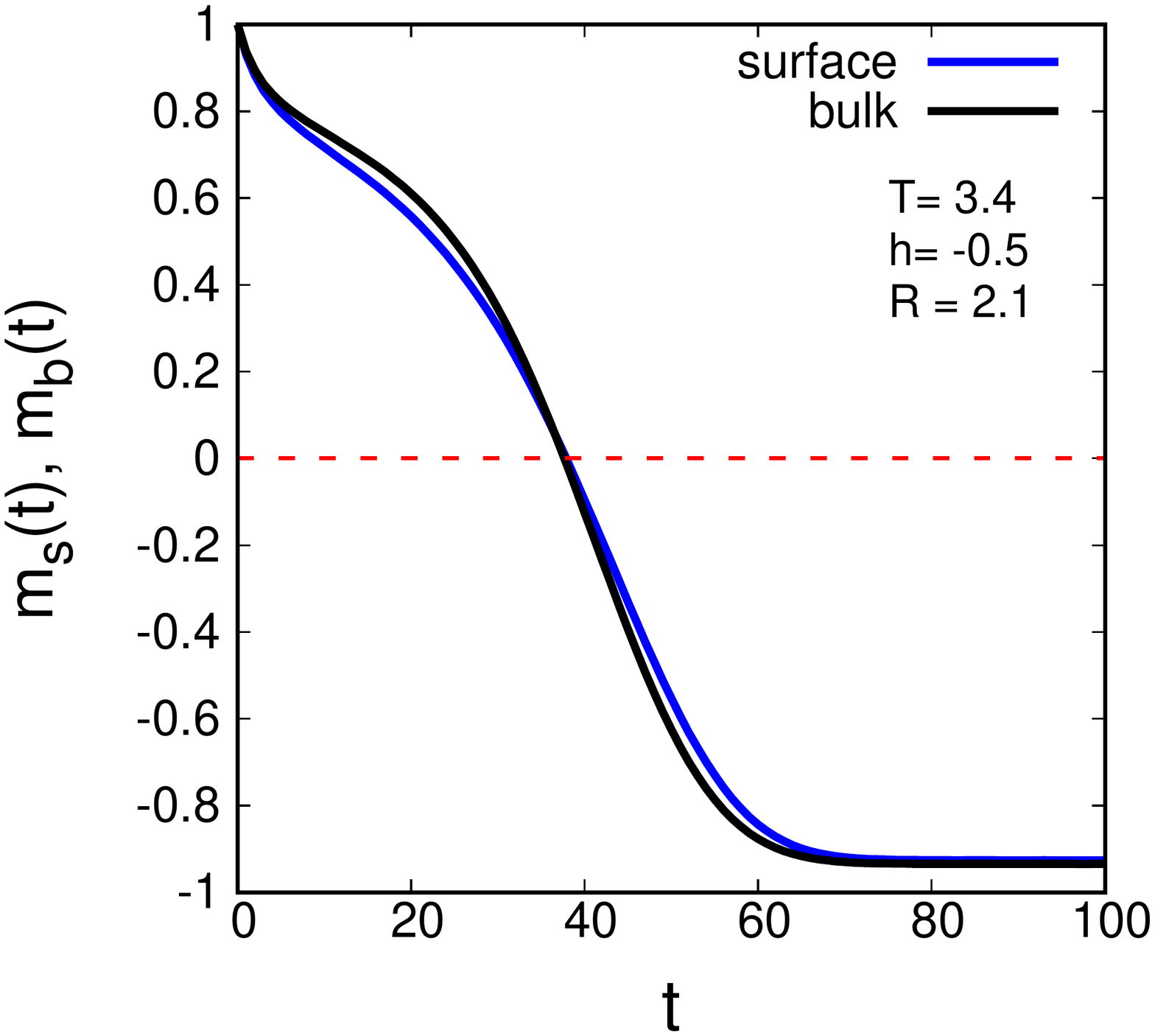}
		\subcaption{}
	\end{subfigure}
	\begin{subfigure}{0.49\textwidth}
		\includegraphics[angle=0,width=1.2\textwidth]{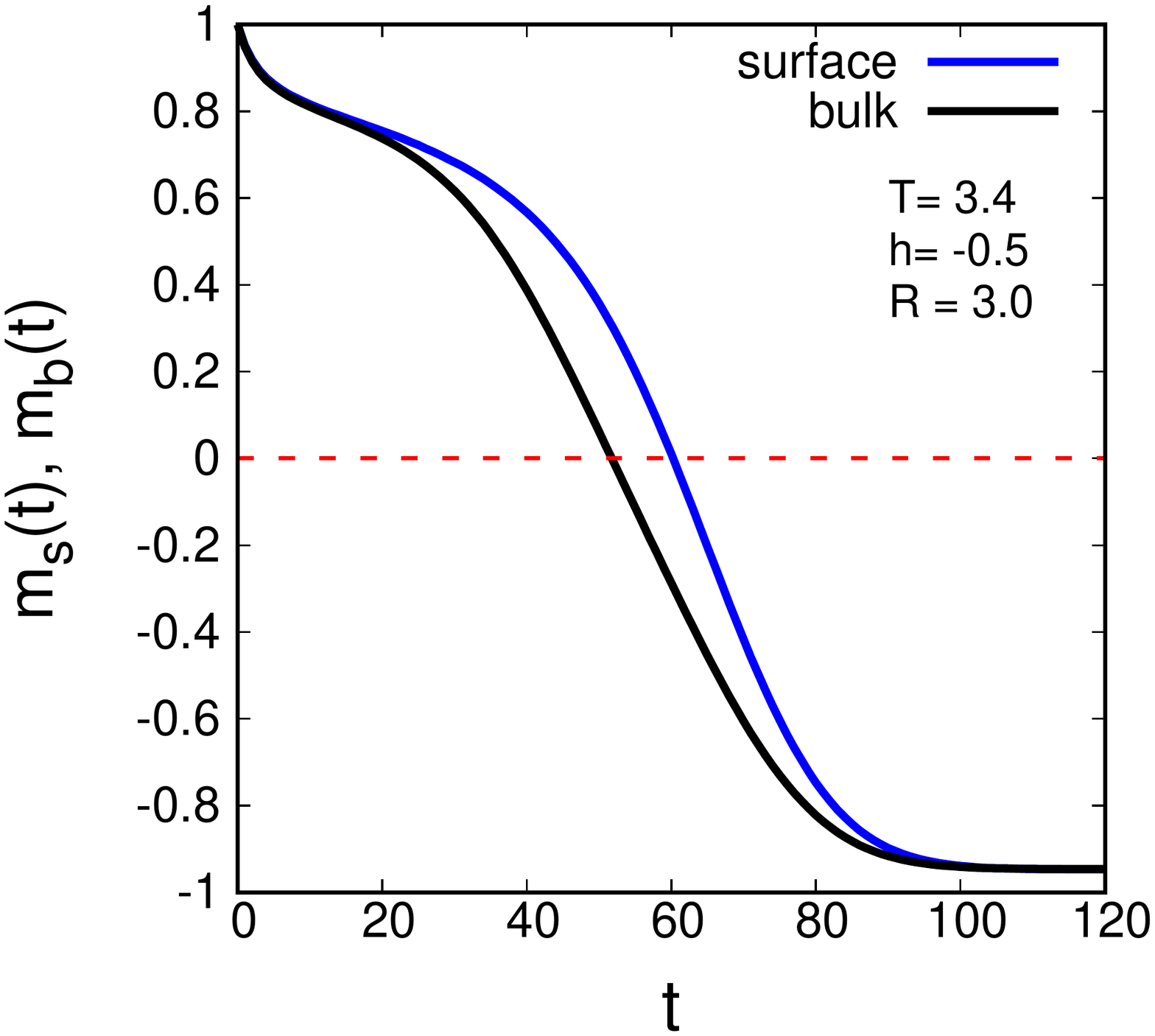}
		\subcaption{}
	\end{subfigure}
	\caption{Variation of magnetisation $ m(t) $ with time $t$ (in unit of MCSS) averaged over 1000 samples for three different relative interaction strengths $ R= J_f/J= 1.2, 2.1, 3.0 $. J is set to $ J=1.0$ always. Temperature is kept fixed at $T=3.4$ (almost $0.75 \; T_c$) and applied field is $h= -0.5$. The size of the system is $L=32$.}
\label{magtime}
\end{figure}

For getting an idea about the reversal time of surface and bulk, evolution of magnetisation with time (averaged over 1000 samples) has been studied (Fig-\ref{magtime}) separately for surface and bulk for different relative interaction strengths $R=J_f/J$ in presence of a negative field. For simplicity, $J$ is set to $J=1$ throughout the study. Interestingly, a competitive behaviour of reversal times of surface and bulk is noticed depending on the value of $R$. For $J_f \sim J$, (Fig-\ref{magtime}a) the magnetisation of the surface gets reversed faster compared to the bulk since fewer nearest neighbours enhance the probability of flipping of a spin on the surface. Now, if the $J_f$ is increased as if the spins on surface are strongly interacted with the nearest neighbours, then it is difficult to flip those spins easily. Obviously then both the surface and bulk reversal times increase. Surprisingly, we found the existence of a certain critical relative interaction strength $R_c=(J_f^c/J)$ for which metastable lifetimes of the surface and the bulk become almost equal (Fig-\ref{magtime}b). And beyond that critical value $R>R_c$, an opposite scenario is observed where reversal of bulk magnetisation becomes faster than that of  surface magnetisation (Fig-\ref{magtime}c). 

\begin{figure}[h!]
	\centering
	\begin{subfigure}{0.49\textwidth}
		\includegraphics[angle=0,width=1.2\textwidth]{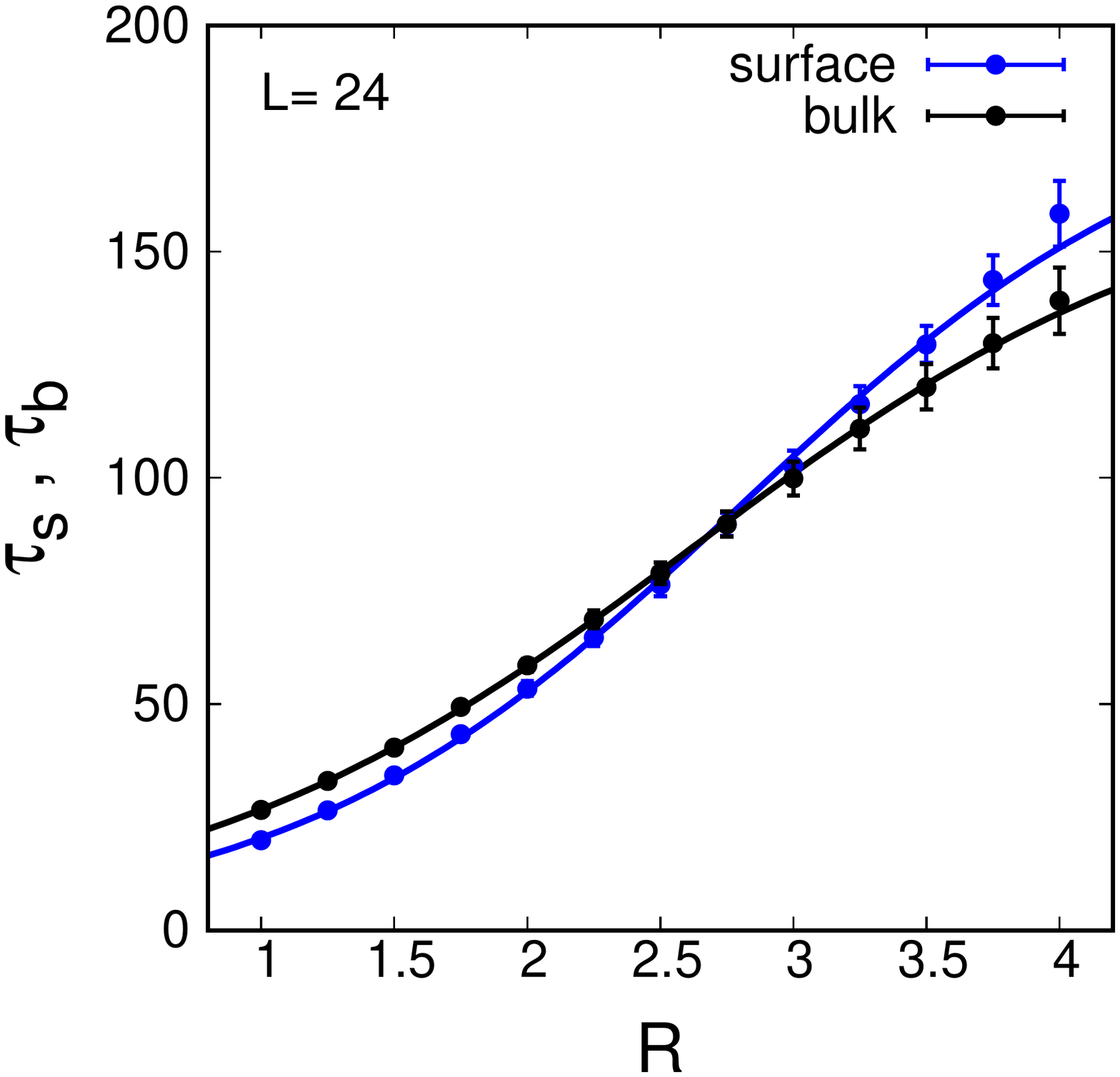}
		\subcaption{}
	\end{subfigure}
	\begin{subfigure}{0.49\textwidth}
		\includegraphics[angle=0,width=1.2\textwidth]{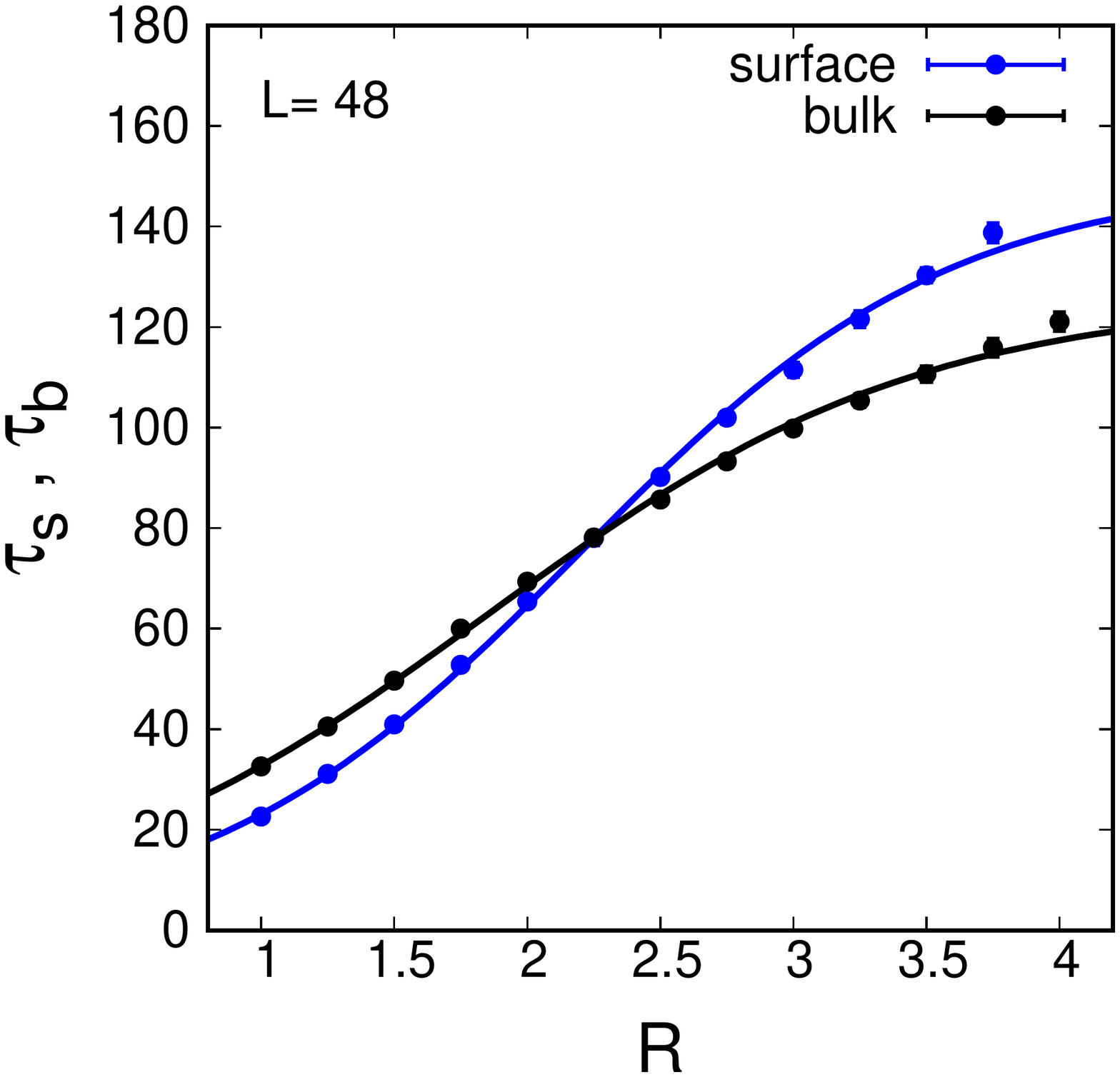}
		\subcaption{}
	\end{subfigure}
	\caption{Mean reversal time of surface $\tau_s$ and bulk $\tau_b$ with the relative interaction strength $ R $ at temperature $T=3.2$ for the system of size $L=24,48$. Applied field is $h= -0.5$. Data are fitted to the function $f(x)= \frac{a}{1+e^{(b-x)/c}}$. Using this function with fitting parameters, critical relative interaction strength $R_c$ is found out with certain radius of convergence for which reversal time of surface becomes equal to that of bulk ($\tau_s \simeq \tau_b $ ).}
	\label{fsize1}
\end{figure}

\begin{figure}[h!]
	\centering
	\includegraphics[angle=0,width=0.6\textwidth]{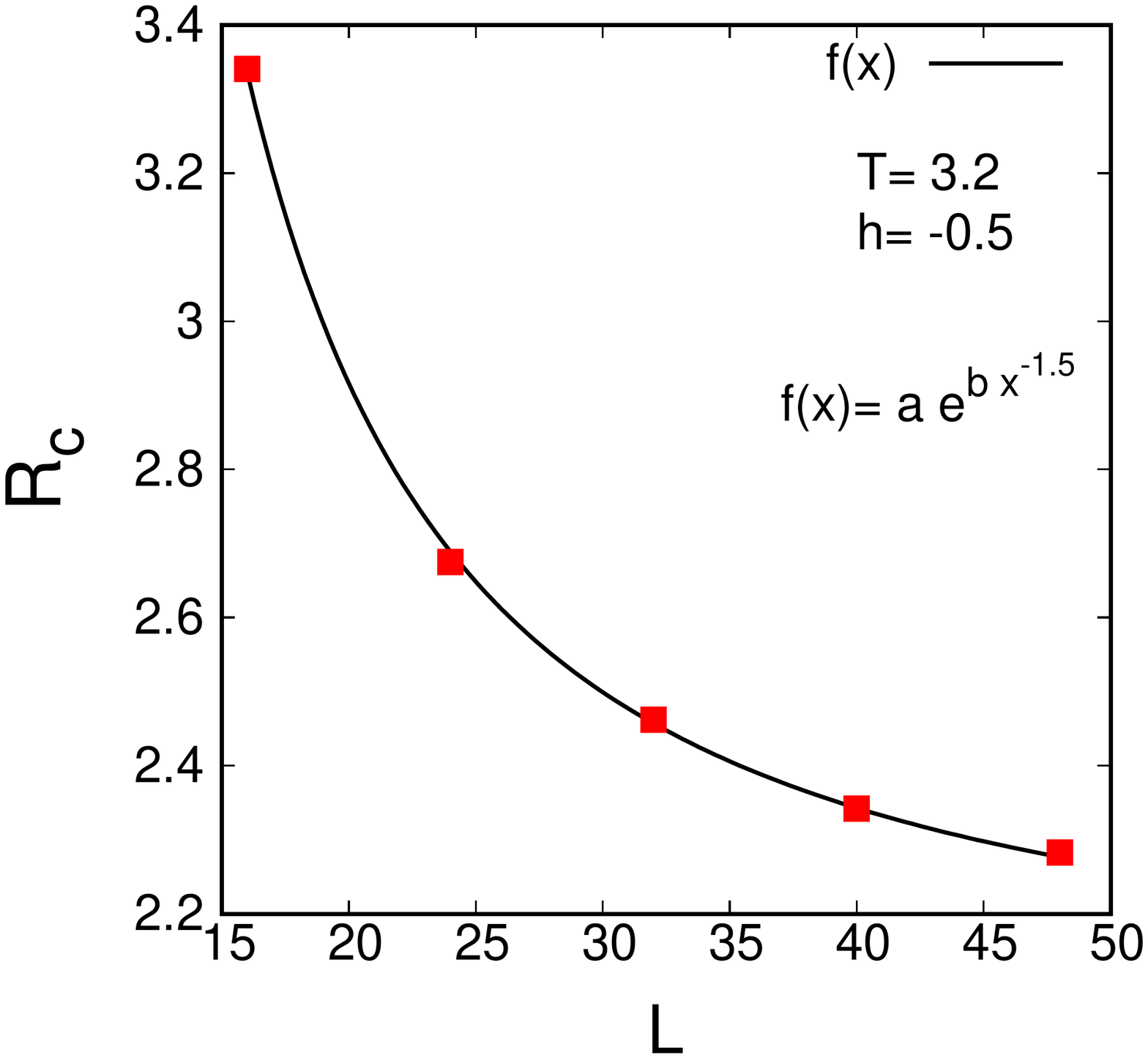}
	\caption{Critical relative interaction strength $R_c$ (for which $\tau_s = \tau_b $ ) for five different sizes of the system $L=16,24,32,40,48$. Data fit to the function $f(x)= a e^{b x^{-1.5}}$ with $a=2.078 \pm 0.007$ and $b=30.3 \pm 0.3$.}
	\label{fsize2}
\end{figure}

Since we are simulating a very small size of lattice we must check the finite size effect. For that, mean reversal times of surface $\tau_s$ and bulk $\tau_b$ have been calculated over 1000 different samples and studied (Fig-\ref{fsize1}) with the variation of relative interaction strength $R$ at a fixed temperature $T=3.2$ and applied field $h=-0.5$ for different sizes of lattice $L=16,24,32,40,48$. The dependencies of $\tau_s$ and $\tau_b$ on $R$ are fitted to the exponential function $f(x)= \frac{a}{1+e^{(b-x)/c}}$. In Fig-\ref{fsize1}, we have presented it for the lattice of sizes $L=24$ and $48$ only. In obvious manner, both the $\tau_s$ and $\tau_b$ increase with the increase of $R$ as the stronger coupling helps to delay the reversal. Errors are evaluated by simple block averaging method and the chi-square goodness of fit test has been accomplished supported by the Q-values provided in Table-\ref{table1}. Now using this function with the fitting parameters (a,b and c), the distinct critical relative interaction strengths $R_c$ have been identified (Fig-\ref{fsize2}) with certain radius of convergence 0.01 (within which two curves intersect) for different sizes of lattice. Clearly, for the critical relative interaction strength $R=R_c$, the surface reversal synchronizes with the bulk reversal ($ \tau_s \simeq \tau_b $), within a value 0.01 of closeness. Variation of $R_c \pm 0.01$ with size of the system $L$ fit (see, Table-\ref{table2}) to the function $f(R_c) \sim a \; e^{(b L^{-1.5})}$ with $a=2.078 \pm 0.007$ and $b=30.3 \pm 0.3$. It is worth mentioning that, in the thermodynamic limit, the value of $R_c$ tries to reach a fixed value $R_c \simeq 2.078$ (Fig-\ref{fsize2}).

\begin{figure}[h!]
\centering
	\begin{subfigure}{0.49\textwidth}
		\includegraphics[angle=0,width=1.2\textwidth]{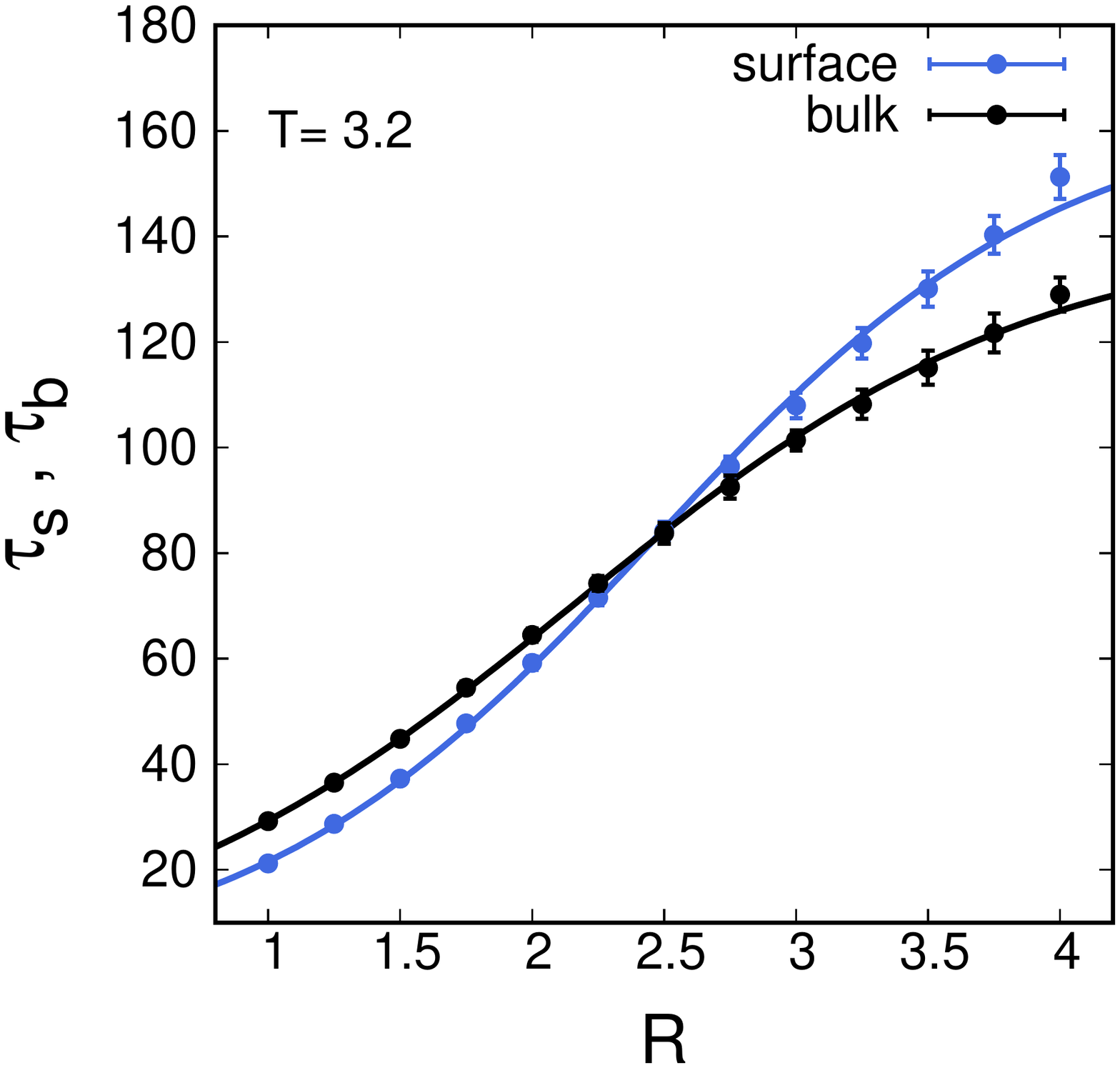}
		\subcaption{}
	\end{subfigure}
	\begin{subfigure}{0.49\textwidth}
		\includegraphics[angle=0,width=1.2\textwidth]{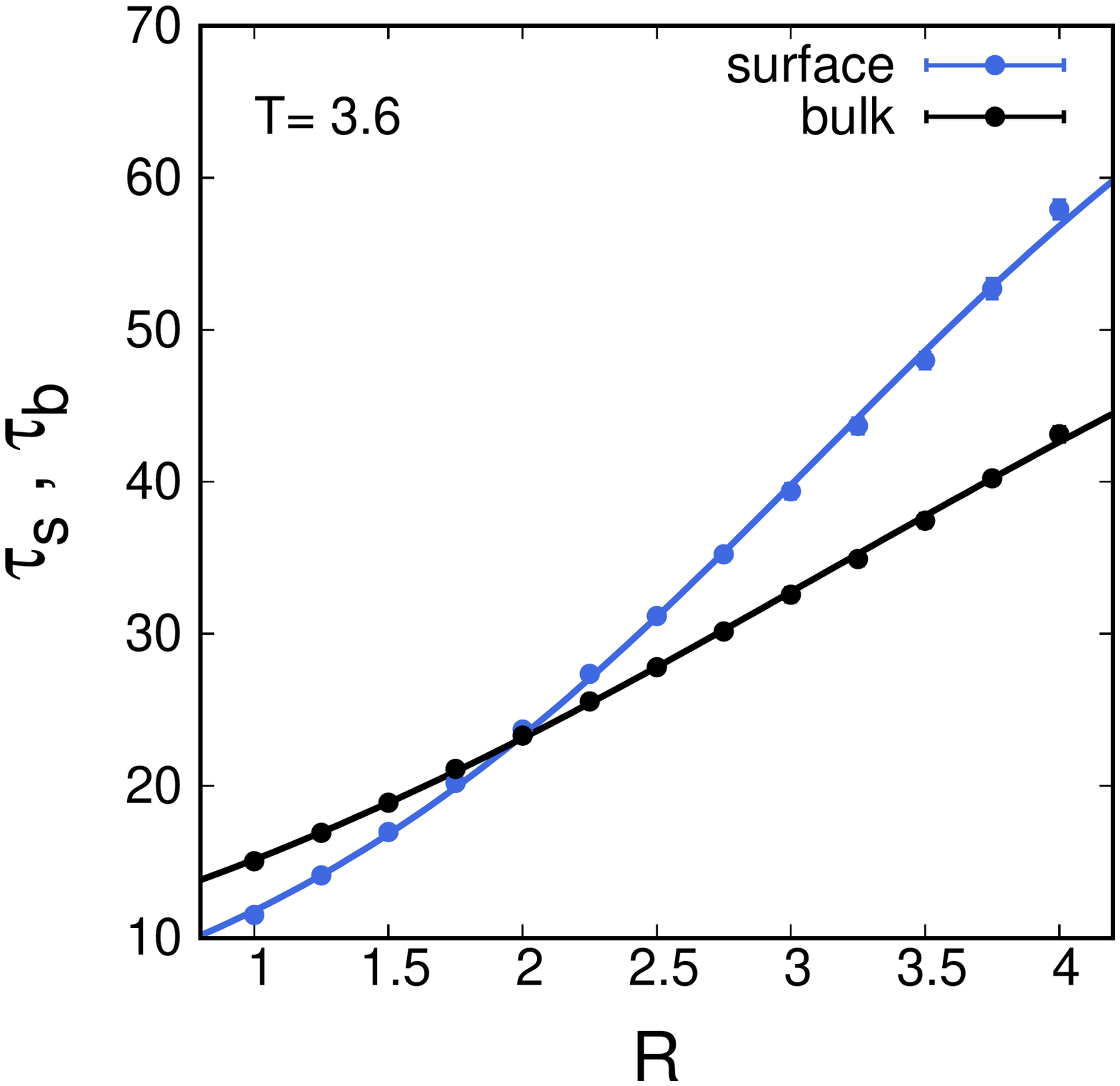}
		\subcaption{}
	\end{subfigure}
	\caption{Mean reversal time of surface $\tau_s$ and bulk $\tau_b$ with the relative interaction strength $R=J_f/J$ at temperature $T=3.2$ and $T=3.6$. Applied field is $h= -0.5$. Data are fitted to the function $f(x)= \frac{a}{1+e^{(b-x)/c}}$. The size of the system is $L=32$. }
	\label{rt_jratio}
\end{figure}

An investigation of the dependencies of reversal times on temperature should be expected. We have studied the variation of mean surface reversal time $\tau_s$ and mean bulk reversal time $\tau_b$ with $R$ for different temperatures in presence of a fixed applied field $h= -0.5$. In Fig-\ref{fsize1} the study is presented for $T=3.2$ and $T=3.6$ along with the errors determined by the block-averaging method. Fitting (see, Table-\ref{table3}) of data is analogous to the Fig-\ref{fsize1}. $R_c$ is found out for each temperature in the same fashion as we did for Fig-\ref{fsize1}. Radius of convergence (here 0.01) of the two fitted functions is considered as the error (of the order of size of the data point) of the determination of $R_c$. The Fig-\ref{rt_jratio} depicts that the critical relative interaction strength $R_c$ is strongly dependent on the temperature.

\begin{figure}[h!]
\centering
	\begin{subfigure}{0.49\textwidth}
		\includegraphics[angle=0,width=1.2\textwidth]{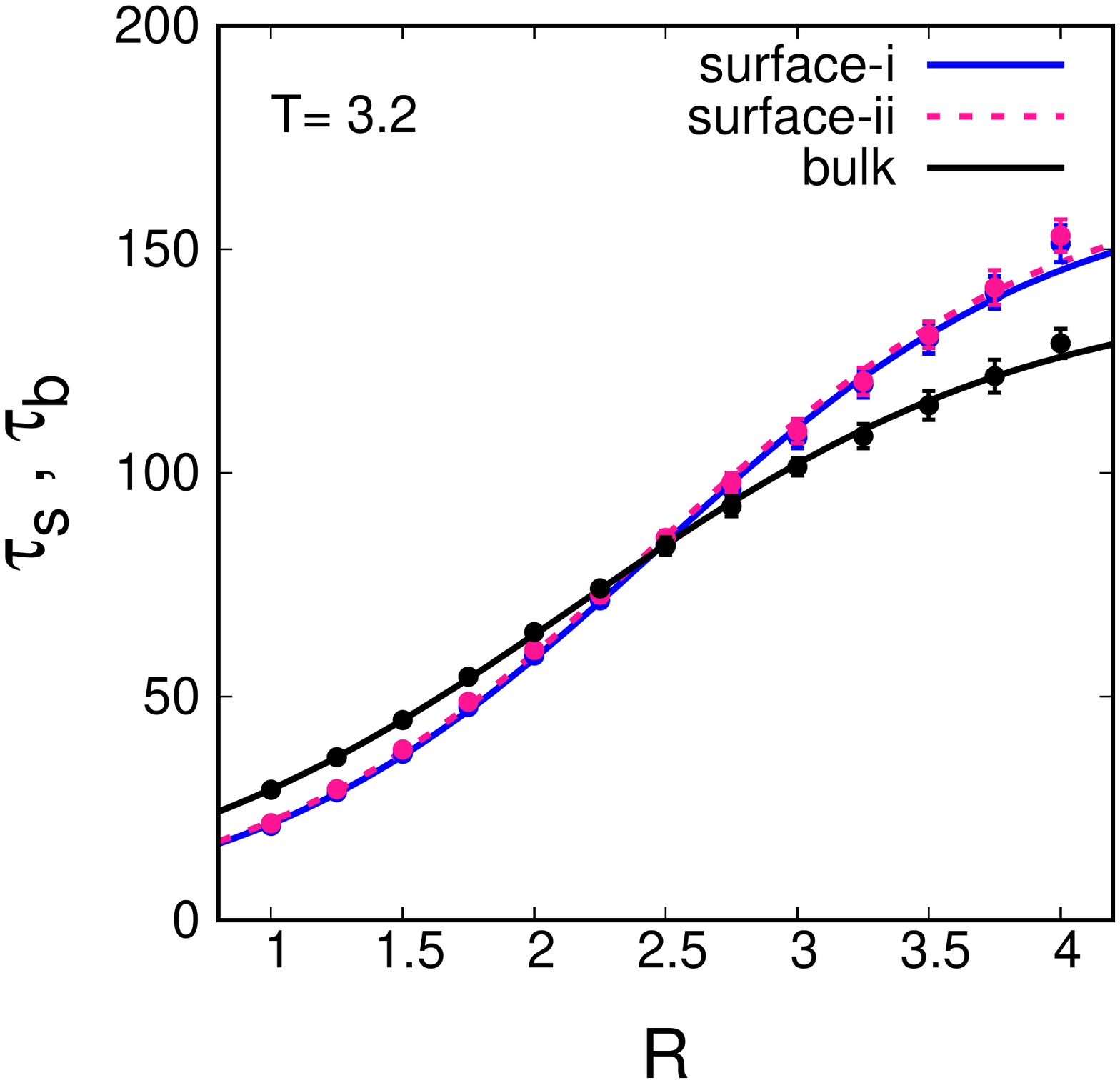}
		\subcaption{}
	\end{subfigure}
	\begin{subfigure}{0.49\textwidth}
		\includegraphics[angle=0,width=1.2\textwidth]{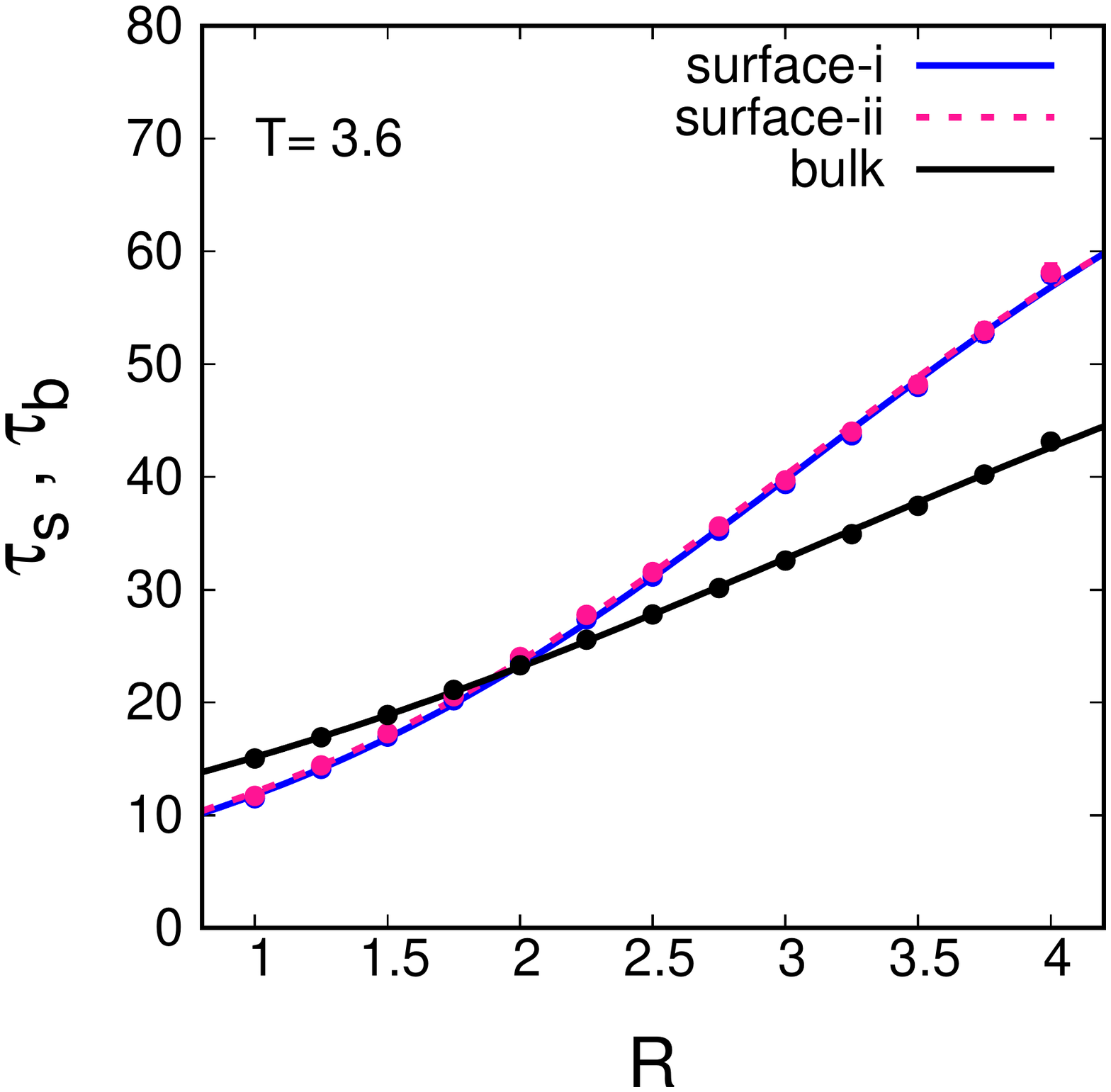}
		\subcaption{}
	\end{subfigure}
	\caption{Same study of Fig-\ref{rt_jratio} has been worked out again with additional pink-coloured dashed line, which indicates the mean reversal time of the {\it restricted-surface} (surface-ii). Here, we have considered $L=32$. }
	\label{rt_jratio2}
\end{figure}

In Fig-(\ref{rt_jratio2}), the study of Fig-(\ref{rt_jratio}) has been repeated with additional consideration of {\it restricted-surface} (so called surface-ii). The pink dashed line indicates the mean reversal time of the {\it restricted-surface} which is defined by excluding the spins on edges and corners. It should be clarified that, the 'surface' in Fig-(\ref{rt_jratio}) and the 'surface-i' in Fig-(\ref{rt_jratio2}) are the same. For detailed description of the surface, we want to drive the reader's attention back to the previous section. From Fig-(\ref{rt_jratio2}) it can be observed that, the deviation of pink line from the blue line is very small. Definitely, for large $L$ limit, that deviation would be vanishingly small. Though the spins on the edges and corners are of different categories, but their negligible impact prompted us to carry out the whole studies by considering the so called surface-i (or, surface).

\begin{figure}[h!]
	\centering
	\includegraphics[angle=0, width= 0.6\textwidth]{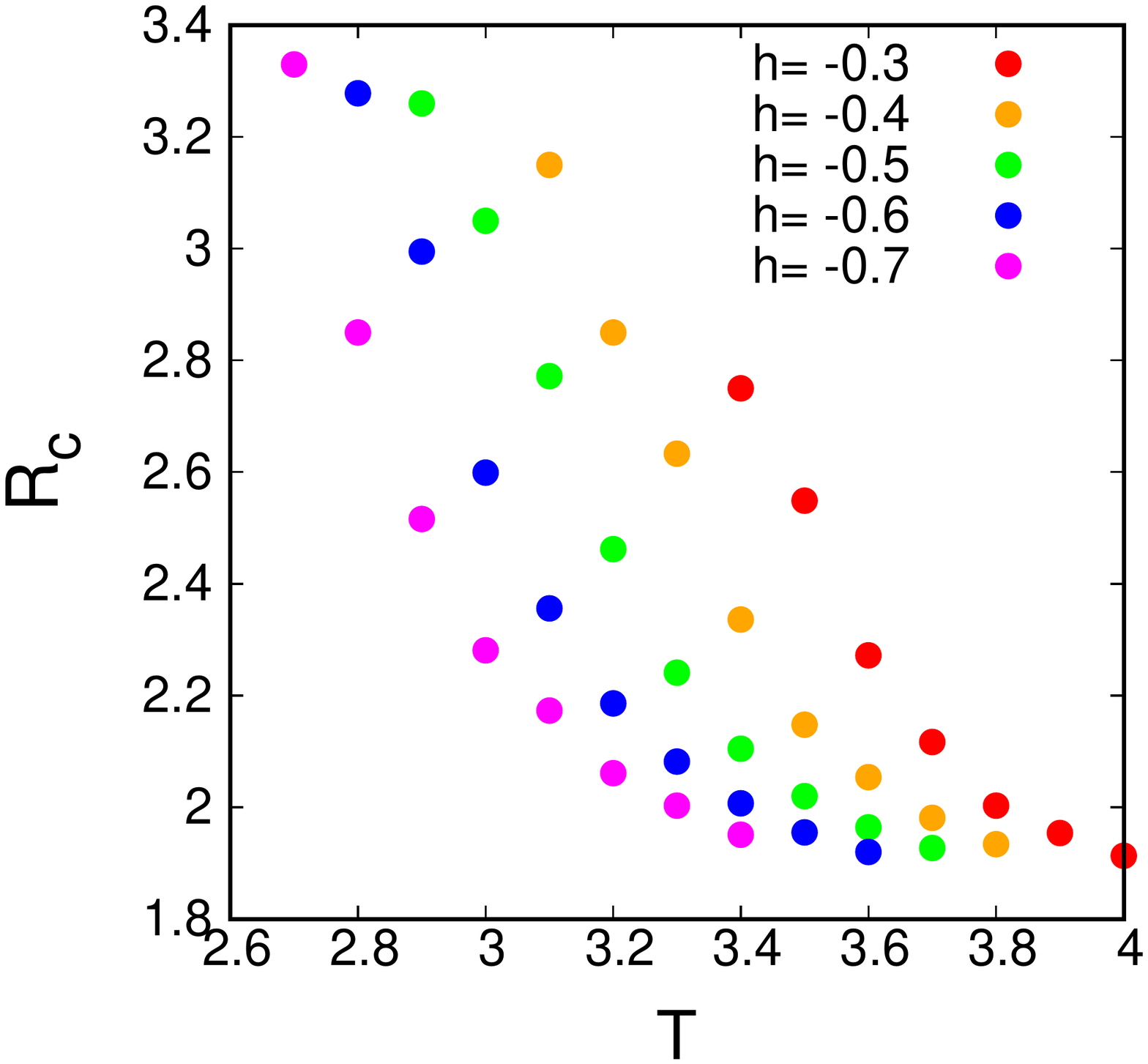}
	\caption{Variation of critical relative interaction strength $R_c$ (for which $\tau_s = \tau_b $ ) with temperature $T$ for five different strengths of applied fields $h= -0.3,-0.4,-0.5,-0.6,-0.7$. The size of the system is $L=32$.}
	\label{critj_t}
	\end{figure}

\begin{figure}[h!]
	\centering
	\includegraphics[angle=0,width=0.6\textwidth]{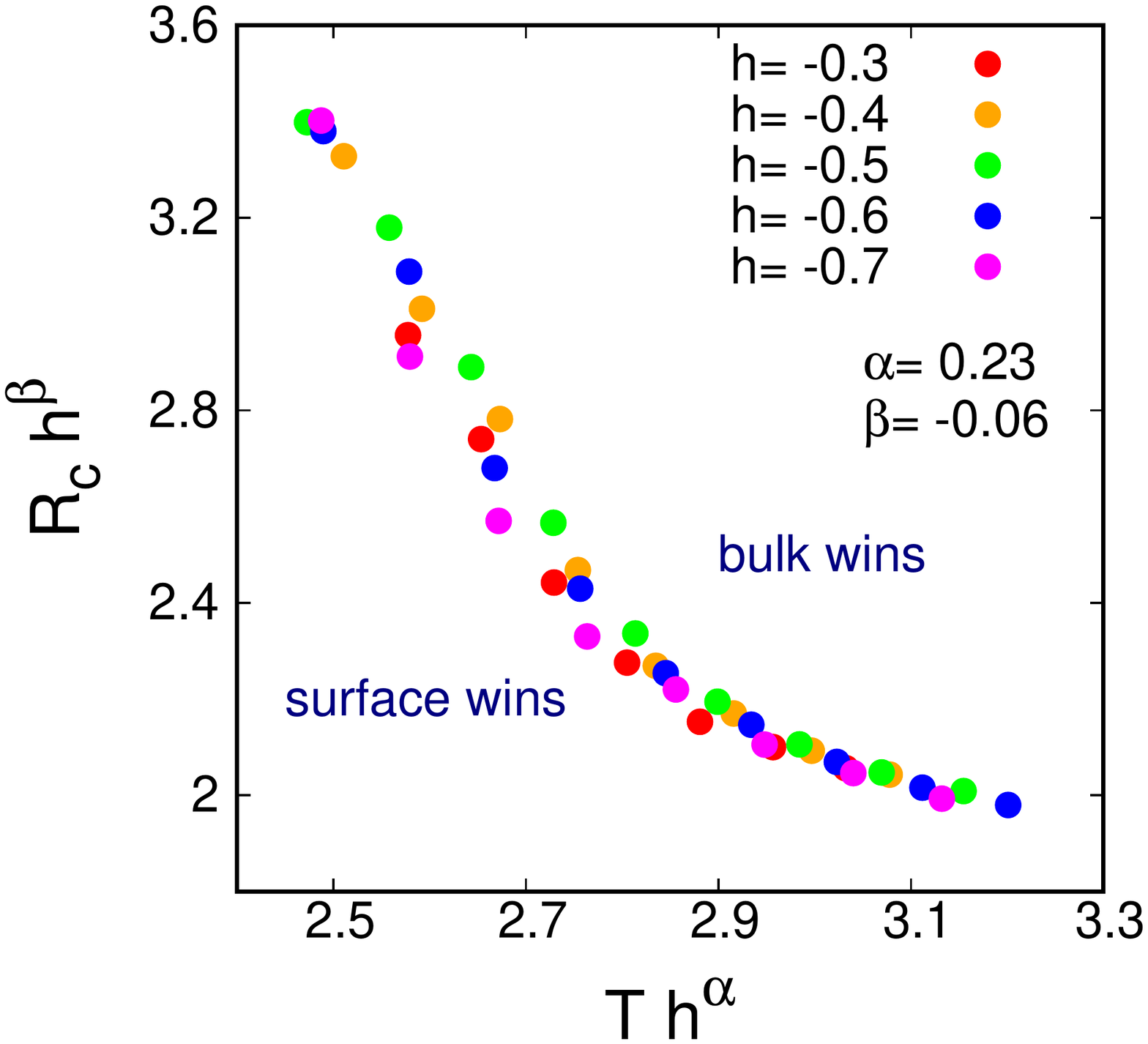}
	\caption{Variation of scaled critical relative interaction strength $R_c h^\beta$ with scaled temperature $T h^\alpha$ with exponents $\alpha= 0.23 \pm 0.01$ and $\beta= -0.06 \pm 0.01$. Collapsed data follow the scaling relation $R_c \sim h^{-\beta} f(Th^\alpha)$. In the upper region, the bulk wins in faster reversal whereas in the lower part the surface wins over the bulk. The size of the system is $L=32$.}
	\label{scaling}
\end{figure}

The dependency of $R_c$ on the temperature has been investigated for different strengths of applied field (Fig-\ref{critj_t}). At a fixed strength of applied field, the $R_c$ is identified within a suitable range of temperature where the difference of $\tau_s$ and $\tau_b$ is prominent. $R_c$ decreases with the increase of temperature and also strength of the applied field. Interestingly, data are collapsed for scaled relative interaction strength $R_c h^\beta$ and scaled temperature $T h^\alpha$ with $\alpha = 0.23 \pm 0.01$ and $\beta = -0.06 \pm 0.01$. Exponents are optimized (visually) by simple trial and error method to get the data collapsed (Fig-\ref{scaling}). So the relative interaction strength ($R$) is found to follow a scaling relation with temperature and applied field, $R_c \sim h^{-\beta} f(Th^\alpha)$. The form of the function $f(Th^\alpha)$ is not yet determined. We want to draw the reader's attention to the interesting fact that, the collapsed data indicates a boundary along which the reversal process of the surface almost synchronizes with that of the bulk. As far as the faster reversal of magnetisation is concerned, below that boundary, the surface-reversal wins over the bulk-reversal. Whereas, above the boundary, the bulk-reversal wins over the surface-reversal.

\vskip 1 cm
\section{Summary}
 
 The metastability of ferromagnetic systems shows various interesting behaviours. The metastable lifetime is measured 
 from the reversal of magnetisation. Out of which, the interesting competitive behaviours between surface-reversal and bulk-reversal have been studied extensively in three dimensional Ising ferromagnet by Monte Carlo simulation using Metropolis  single spin flip algorithm. The bounding layers
 of the simple cubic lattice are considered as surface. The whole system is kept in open boundary conditions in all directions. The surface
 interacts with the core (excluding surface) by a relative interaction strength ($R$). 
 
 For a fixed value of the temperature ($T$) and applied magnetic field ($h$), both the surface and the bulk reversal times grow with the relative interaction strength as $f(x)= \frac {a}{1+e^{\frac{(b-x)}{c}}}$ (Fig.-\ref {rt_jratio}).
 It was observed further that the surface reversal occurs earlier for weak 
 interaction and vice versa. Interestingly, a critical value 
 ($R_c$) of
 relative interaction strength was found, for which both (surface and bulk) reversal occur simultaneously. This critical relative  interaction strength, $R_c$, was found to be a function of temperature ($T$) and the applied magnetic field ($h$). For fixed, $h$, the $R_c$ is a monotonically decreasing function of the temperature $T$. However, an interesting scaling behaviour, like $R_c \sim h^{-\beta}f(Th^{\alpha})$, was found by the method of data collapse. The $f$ is the scaling function and the estimated exponents are $\alpha=0.23\pm0.01$ and
 $\beta=-0.06\pm0.01$ (Fig.-\ref{scaling}).
 
 The critical relative interaction strength $R_c$ decreases monotonically with the system size $L$, for fixed values of the temperature ($T$) and applied magnetic field ($h$). Here, the
 data have been fitted with $R_c \sim e^{bx^{-1.5}}$ (Fig-\ref{fsize2}).
 
 The decay of metastable volume fraction was also observed to reflect the surface and bulk metastability through Avrami's law. Avrami's law depicts that the metastable volume fraction would
 decay at $e^{-(t/{\tau})^{d+1}}$ in $d$-dimensional system. In
 this present study, the metastable surface volume fraction was
 found to decay at $e^{-(t/{\tau})^{3}}$ and that for bulk $e^{-(t/{\tau})^{4}}$ (Fig.-\ref{avrami}). These results are convincing that the bulk and surface have different metastable behaviours.
 
 In short, the surface and bulk metastable behaviours show various interesting effects, in Ising ferromagnet, which show further scopes of  research. Mainly the reversal time can be suitably varied by changing the relative interaction strength. Although the real materials hardly follow the Metropolis dynamics, the results presented here, gives an idea of competition between the reversal times
 for bulk and surface, at least in the model system.

\section {Acknowledgements}

MN would like to acknowledge Swami Vivekananda Scholarship (SVMCMS) for financial 
support. MA acknowledges FRPDF grant of Presidency University for financial support.



\newpage
\newpage

\newpage
\newpage

\newpage

\newpage

\newpage


\newpage

\newpage



\begin{table}[ht!] 
	\caption{Fitting parameters of Fig-\ref{fsize1}}
	\centering
	\begin{tabularx}{0.7\textwidth} { 
			| >{\centering\arraybackslash}X 
			| >{\centering\arraybackslash}X
			| >{\centering\arraybackslash}X
			| >{\centering\arraybackslash}X 
			| >{\centering\arraybackslash}X | }
		\hline
		L & surface/bulk & $ \chi^2 $ & DOF & Q\\
		\hline	
		24 & surface & 4.597 & 10 & 0.92 \\
		24 & bulk & 0.526 & 10 & 0.99 \\
		48 & surface & 13.79 & 9 & 0.13 \\
		48 & bulk & 11.79 & 10 & 0.30 \\		
		\hline		
	\end{tabularx}
	\label{table1}
\end{table}

\begin{table}[ht!] 
	\caption{Fitting parameters of Fig-\ref{fsize2}}
	\centering
	\begin{tabularx}{0.4\textwidth} { 
				| >{\centering\arraybackslash}X 
				| >{\centering\arraybackslash}X
				| >{\centering\arraybackslash}X | }
		\hline
		$ \chi^2 $ & DOF & Q\\
	    \hline	
		2.881 & 3 & 0.41 \\		
		\hline		
	\end{tabularx}
	\label{table2}
\end{table}

\begin{table}[ht!] 
	\caption{Fitting parameters of Fig-\ref{rt_jratio}}
	\centering
	\begin{tabularx}{0.7\textwidth} { 
		| >{\centering\arraybackslash}X 
		| >{\centering\arraybackslash}X
		| >{\centering\arraybackslash}X
		| >{\centering\arraybackslash}X 
		| >{\centering\arraybackslash}X | }
	\hline
	T & surface/bulk & $ \chi^2 $ & DOF & Q\\
	\hline	
		3.2 & surface & 6.814 & 10 & 0.74 \\
		3.2 & bulk & 2.017 & 10 & 0.99 \\
		3.6 & surface & 13.18 & 10 & 0.21 \\
		3.6 & bulk & 4.183 & 10 & 0.93 \\		
		\hline		
	\end{tabularx}
	\label{table3}
\end{table}

\end{document}